\documentclass[aps,prd,superscriptaddress,showpacs,amsmath,amssymb,twocolumn]{revtex4}
\usepackage{graphicx,epsf}
\usepackage[]{latexsym}
\newcommand{\be}{\begin{eqnarray}}
\newcommand{\ee}{\end{eqnarray}}
\newcommand{\bra}[1]{\mbox{$\langle\, #1 \mid$}}

\newcommand{\ket}[1]{\mbox{$\mid #1\,\rangle$}}

\newcommand{\pro}[2]{\mbox{$\langle\, #1 \mid #2\,\rangle$}}
\newcommand{\expec}[1]{\mbox{$\langle\, #1\,\rangle$}}

\renewcommand{\d}{\mbox{{\rm d}}}
\begin{document}
\title{Boundaries and the Casimir effect in non-commutative space-time}
\author{R.~Casadio}
\email{casadio@bo.infn.it}
\affiliation{Dipartimento di Fisica, Universit\`a di Bologna,
via Irnerio 46, 40126 Bologna, Italy}
\affiliation{I.N.F.N., Sezione di Bologna, via Irnerio 46, 40126 Bologna, Italy}
\author{A.~Gruppuso}
\email{gruppuso@iasfbo.inaf.it}
\affiliation{I.N.A.F.-I.A.S.F. Bologna, via Gobetti 101, 40129 Bologna, Italy}
\affiliation{I.N.F.N., Sezione di Bologna, via Irnerio 46, 40126 Bologna, Italy}
\author{B.~Harms}
\email{bharms@bama.ua.edu}
\affiliation{Department of Physics and Astronomy, The University
of Alabama, Box 870324, Tuscaloosa, AL 35487-0324, USA}
\author{O.~Micu}
\email{micu001@bama.ua.edu}
\affiliation{Department of Physics and Astronomy, The University
of Alabama, Box 870324, Tuscaloosa, AL 35487-0324, USA}
\begin{abstract}
We calculate modifications to the scalar Casimir force between two
parallel plates due to space-time non-commutativity. We devise a
heuristic approach to overcome the difficulties of describing
boundaries in non-commutative theories and predict that boundary
corrections are of the same order as non-commutative volume
corrections. Further, both corrections have the form of more
conventional finite surface effects.
\end{abstract}
\maketitle
\large
\section{Introduction}
\label{intro}
The idea that space-time coordinates do not commute below a certain
length scale is a rather old one~\cite{snyder}.
As such, it has been recently revived in string theory~\cite{witten,seiberg},
where one deals with spatially extended objects, and in the search
for a quantum theory of gravity~\cite{moffat} where the Planck length
naturally appears.
Non-commutativity itself is a common feature of quantum theories.
It is manifested in quantum mechanics in the phase-space commutation
relations
\be
\left[p_i,x_j\right]= i\,\hbar\,\delta_{ij}
\ ,
\ee
and in quantum field theory in the commutation relations of creation
and annihilation operators.
Also Yang-Mills theories on non-commutative spaces~\cite{witten1}
appear in string theory and M-theory.
\par
Non-commutative geometry is based on the concept that there might
exist a fundamental length (which we shall always denote as
$\ell$) in the fabric of space-time~\cite{snyder}. For a parameter
to be considered a fundamental length, it should respect Lorentz
invariance, and therefore the time coordinate needs to be included
among the non-commutative variables. However, this is not a
trivial change and theories in which the time coordinate does not
commute with spatial ones seem to be acausal. An example of such a
theory is given in Ref.~\cite{seiberg}, where the authors study
the effects of space-time non-commutativity on the scattering of
wave packets and show that stringy effects cancel the acausal
effects that appear in field theories. Other motivations for
non-commutativity come from D-brane
scenarios~\cite{witten,seiberg} where space non-commutativity
appears when D-branes occur in hyper-magnetic fields, and time
non-commutativity is generated similarly by nonzero hyper-electric
fields~\footnote{From the experimental point of view, there is no
compelling evidence that favors non-commutativity. We however
mention that non-commutative quantum electrodynamics~\cite{ncqed}
might explain the results found by the PVLAS
experiment~\cite{pvlas}.}.
\par
Non-commutativity in space-time is usually implemented by
replacing the ordinary product with the Moyal
$\star$-product~\cite{moyal}.
One then finds that the free propagator is not changed, and a
(self-)interaction in the field theory is necessary in order to see
any corrections with respect to the commutative case
(for a recent review, see {\em e.g.}, Ref.~\cite{wohlgenannt}).
An alternative, coherent state approach was proposed in
Ref.~\cite{smailagic} which leads to non-trivial corrections
already at the level of free propagators, as one would more
naturally expect (for some of its consequences, see
Refs.~\cite{cons,mirror}).
\par
The Casimir force between perfectly conducting (parallel) neutral
plates per unit area~\cite{casimir}
(we shall use natural units with $c=\hbar=1$),
\be
\mathcal{F}_0=\frac{F_0}{\mathcal{A}} =
-\frac{\pi^2}{240\,L^4}
\ ,
\label{Ccom}
\ee
where $\mathcal{A}$
is the area of the plates and $L$ the separation between the
plates ($\mathcal{A}\gg L^2$), is perhaps the most prominent
example of non-trivial vacuum effects in quantum electrodynamics.
It has been studied extensively and can be measured with very high
accuracy (for a review, see Refs.~\cite{bordag,milton}).
Several papers have appeared in the literature which consider the
corrections due to the existence of a minimal
length~\cite{hossenfelder1} or of compactified extra spatial
dimensions~\cite{hossenfelder2}. Because of the high accuracy of
the measurements, the Casimir effect is also a natural candidate
to show possible effects due to short-distance space-time
non-commutativity~\cite{nam}.
It is important to note that the
final result in general depends on how non-commutativity is
realized. For example, in the usual Moyal approach~\cite{moyal},
non-commutative corrections will depend on the details of the
(self)interaction terms which are included in the field
theory~\cite{nam}. In the approach of Ref.~\cite{smailagic}, which
we shall follow, such corrections should not depend on coupling
constants and therefore look universal~\footnote{This property
will be fully preserved only for volume corrections, whereas
boundary corrections will depend on the coupling between the field
and the plates.}.
\par
In describing the Casimir force, one usually represents the plates
as Dirac $\delta$-function sources or, equivalently, as boundary
conditions on the field modes.
Thus, a crucial point which needs to be addressed before
introducing non-commutativity in the
Casimir effect is how boundaries are described in a
non-commutative space-time.
>From a physical point of view, it is meaningless to assign a position
with accuracy better than the minimum length admitted in the theory.
Therefore what should be done is to give up the boundary conditions
and, for example, describe the plates as sources ``smeared'' over a
size of the order of the fundamental length $\ell$, making sure at the same
time that the necessary conditions for having a discrete spectrum
of modes between the plates continue to hold.
Unfortunately, such prescriptions do not uniquely fix the interaction
term in the Lagrangian which describes the plates explicitly, and it is
hopeless to obtain exact expressions for the non-commutative modes
given non-trivial potentials for the plates.
\par
Based on the above observation, we shall therefore attempt a more
modest approach by treating the non-commutativity around the
boundaries ``perturbatively'' with respect to the standard treatment.
This will allow us to show to what extent approximate
calculations which make use of the usual boundary conditions can
still be employed and find that, to leading order in the ratio
between the non-commutative length and the plate separation,
boundary corrections are expected to be of the same order as
volume corrections and both corrections have the form of finite
surface effects.
\par
In Section~\ref{NC_st}, we shall review the coherent state
approach to quantum  field theory in non-commutative space-time
and then compute non-commutative corrections to the Casimir effect
in Section~\ref{NC}. We shall begin by discussing the problem of
defining boundaries in a non-commutative space-time in
Section~\ref{NC_b} where we estimate boundary effects and proceed
with volume corrections in Section~\ref{NC_v}. The comparison with
surface effects in the standard description will finally be
discussed in Section~\ref{surf} and more conclusions drawn in
Section~\ref{conc}.
\section{non-commutative space-times}
\label{NC_st}
A $D$-dimensional non-commutative space-time can be defined
in terms of space-time coordinates $x^\mu$ (where $\mu=1,2,\ldots,D$)
which satisfy the following commutation relations:
\be
\left[x^\mu,x^\nu\right]=i\,\Theta^{\mu\nu}
\ ,
\ee
where $\Theta^{\mu\nu}$ must be an antisymmetric Lorentz
tensor~\cite{smailagic}.
As such, it can be transformed into a block-diagonal form:
\be
\hat\Theta^{\mu\nu}={\rm diag}
\left(\hat\Theta_1,\hat\Theta_2,\ldots,\hat\Theta_{D/2}\right)
\ ,
\ee
where
\be
\hat\Theta_i=\Theta_i\, \left( {\begin{array}{cc}
0 & 1
\\
[2ex] -1 &  0
\\
\end{array}}
\right)
\ .
\label{tens}
\ee
Since a coordinate reversal changes the sign of one of these $\Theta_i$,
we can without loss of generality require that all $\Theta_i$ be positive.
\par
In order to have full non-commutativity, one needs to work in a
space-time that has an even number of dimensions.
Then the $D=2\,d$ coordinates can be represented by $d$ two-vectors:
\be
\hat{x}^\mu&=&\left(\hat{x}^1,\hat{x}^2,\ldots,
\hat{x}^{2d-1},\hat{x}^{2d}\right)
\nonumber
\\
&=& \left(\vec{\hat{x}}_1,\vec{\hat{x}}_2,\ldots,
\vec{\hat{x}}_{d}\right)
\ ,
\ee
with $\vec{\hat{x}}_{i} \equiv \left(\hat{y}_{1i},\hat{y}_{2i}\right)$
being two-vectors in the $i$-th non-commutative plane that satisfy
\be
\left[\hat{y}_{1i},\hat{y}_{2i}\right]=i\,\Theta_i
\ .
\label{commutation}
\ee
\par
In a coherent state approach a set of commuting ladder operators
is constructed from non-commutative space-time coordinates
only~\cite{smailagic}.
We define the ladder operators for the $i$-th plane in the following way,
\be
\begin{array}{l}
\hat a_i=\frac{1}{\sqrt{2}}\left(\hat y_{1i}
+i\,\hat y_{2i}\right)
\\
\\
\hat a_i^{\dagger}=\frac{1}{\sqrt{2}}\left(\hat y_{1i}-i\,\hat y_{2i}\right)
\ .
\end{array}
\ee
These operators will then satisfy the canonical commutation rules
\be
\left[\hat a_i,\hat a_j^{\dagger}\right]=\delta_{ij}\,\Theta_i
\ .
\ee
Normalized ($\pro{\alpha}{ \alpha}=1$) coherent states can now be defined
for these operators as
\be
\ket{\alpha}=
\prod_{i}\,\exp\left[\frac{1}{\Theta_i}\,
\left(
\overline{\alpha}_i\,\hat a_i-\alpha_i\,\hat a_i^{\dagger}\right)\right]
\ket{0}
\ ,
\label{co_state}
\ee
where $\ket{0}$ is the vacuum annihilated by all $\hat a_i$.
\par
The usual commutative coordinates are associated with the non-commuting
ones as their mean values over coherent states.
In this way, a non-commutative plane wave can be calculated using Hausdorff
decomposition, resulting in the following form
\be
&&
\bra{\alpha} \exp\left[i\,\sum_{i=1}^{d}\,
\left(\vec{p}\cdot\vec{\hat{x}}\right)_i\right] \ket{\alpha}
\label{mode}
\\
&&
=
\exp\left\{-\sum_{i=1}^{d}\left[ \frac{\Theta_i}{4}\,
(p_{1i}^{2} +p_{2i}^{2})+i\left(\vec{p}\cdot\vec{{x}}\right)_i
\right] \right\}
\ ,
\nonumber
\ee
where $p_{1i}$ and $p_{2i}$ are the momenta canonically conjugate
to the space-time coordinates.
Eq.~(\ref{mode}) then shows that a plane wave in the non-commutative case
will have damping factors proportional to $\Theta_i$ and that we recover the
usual form in the limit $\Theta_i\to 0$.
In a four-dimensional non-commutative space-time we have two such
non-commutative planes ($d=2$) and it is worth noting that, for complete space-time
non-commutativity and $\Theta_i>0$,
the damping factors are positive both for time and space momentum
components, regardless if a Minkowski or Euclidean metric is used.
\par
In the following, for simplicity, we will assume that the parameters which describe
non-commutativity in the $i$-th plane in Eq.~(\ref{commutation}) are
all the same, that is $\Theta_i=\Theta>0$, and denote with
$\ell=\sqrt{\Theta}$ the natural length associated with non-commutativity.
\section{Non-commutative Casimir effect}
\label{NC}
We shall now calculate modifications to the scalar Casimir force
due to the space-time non-commutativity described in the previous
Section.
Following Ref.~\cite{casimir}, we shall do so by
computing the vacuum energy density of a massless scalar field for
the case of two perfectly conducting parallel plates separated by
a distance $L$.
Although in Casimir's original paper, the calculated Casimir force is
due to the vacuum energy density of the electromagnetic field,
several calculations have been carried out using scalar fields satisfying
either Dirichlet or Neumann boundary conditions.
For an extensive review on the subject, see Ref.~\cite{milton}.
\par
The first issue we need to address is how to define the plates in a
non-commutative theory, which will lead us to estimate ``boundary''
corrections due to the ``fuzziness'' of the plates.
Subsequently, we shall also determine volume corrections which are
present in the limit of sharply defined plates.
\subsection{Boundary corrections}
\label{NC_b}
Let us begin by briefly reviewing how boundaries are taken into
account in the standard commutative case.
For this purpose, we consider the simple case of a (complex)
massless scalar field $\phi=\phi(t,x)$ in two-dimensional Minkowski
space-time which interacts with two $\delta$-function potentials,
at $x=0$ and $x=L$.
The dynamics of this system is described by the usual
Lagrangian density
\be
\mathcal{L}=\frac{1}{2}\left(\left|\partial_t\phi\right|^2
-\left|\partial_x\phi\right|^2 \right) +\mathcal{L}_{\rm int}
\ ,
\ee
with
\be \mathcal{L}_{\rm int}=
-\frac{\lambda}{2\,L}\,\delta(x)\left|\phi\right|^2
-\frac{\lambda}{2\,L}\, \delta(x-L)\left|\phi\right|^2
\ ,
\label{oneplusonelag}
\ee
where the coupling constant $\lambda$ is
dimensionless~\cite{milton}.
Integrating around $x=0$ or $x=L$ one
imposes the usual boundary conditions in which the jump of the
first derivative of the field is related to the value of the field
itself at the same point,
\be
\Delta \left( \partial_x \phi\right)|_y
=\frac{\lambda}{L}\,\phi(t,y)
\ ,
\label{bat0andL}
\ee
where $y=0$ or $L$,
\be
\Delta \left(\partial_x\phi\right)|_y=
\partial_x \phi|_{y+\epsilon} -\partial_x \phi|_{y-\epsilon}
\ ,
\ee
and the limit $\epsilon \to 0$ is understood. If we now
take the coupling to infinity, Eq.~(\ref{bat0andL}), together with
regularity of $\phi$ around $x=y$, enforces Dirichlet boundary
conditions at the two points,
\be
\lim_{\lambda \to \infty} \phi(t,y)=0
\ .
\label{bstrongc}
\ee
Upon introducing the usual Klein-Gordon scalar product
\be
\left(\phi_n,\phi_m\right)
=\frac{i}{2}\,\int \d
x\,\left(\phi_n^*\,\partial_t\phi_m-\phi_m\,\partial_t\phi_n^*\right)
\ ,
\ee
one then finds the normal modes
\be
\phi_n(t,x)=
\sqrt{\frac{2}{n\,\pi}}\,e^{-i\,E_n\,t}\,\sin\left(\frac{n\,\pi\,x}{L}\right)
\ ,
\label{phi_n}
\ee
with energy eigenvalues
\be
E_n=\frac{n\,\pi}{L}
\ ,
\ee where $n$ is a positive integer.
\par
When non-commutativity is taken into account, the
$\delta$-functions describing the positions of the plates, should
be replaced by smooth functions which differ from $0$ in an
interval of the order of the minimal length $\ell=\sqrt{\Theta}$
induced by non-commutativity.
At the same time, we require these
potential functions to preserve a {\em discrete spectrum\/} in
between the plates.
To achieve this, for instance, we can {\em superimpose\/} on
each $\delta$-function a ``box'' potential,
\be
g(x)=\left\{
\begin{array}{ll}
\ell^{-1} & {\rm for}\ |x|<\ell
\\
&
\\
0 & {\rm for}\ |x|>\ell \ .
\end{array}
\right.
\label{g}
\ee
The interaction term~({\ref{oneplusonelag})
will then be supplemented by
\be
\mathcal{L}'_{\rm int}=
-\frac{\lambda'}{2\,L}\, g(x)\left|\phi\right|^2
-\frac{\lambda'}{2\,L}\, g(x-L)\left|\phi\right|^2
\ ,
\label{L2}
\ee
where $\lambda'$ is also dimensionless~\footnote{Note that this
non-commutative term becomes of the same form as
$\mathcal{L}_{\rm int}$ in the limit $\ell \to 0$ for all values of
$\lambda'$, and one recovers the commutative theory with
the effective coupling $\lambda\to \lambda+\lambda'$.}.
Eq.~(\ref{bat0andL}) will correspondingly be replaced by
\be
\Delta \left( \partial_x \phi \right)|_{y}
=
\frac{\lambda}{L}\,\phi(t,y) +
\frac{\lambda'}{L\,\ell}\,
\int_{y-\ell}^{y+\ell} \d x \, \phi(t,x)
\ .
\label{NCbatxp}
\ee
In the strong coupling regime
$\lambda\to\infty$, regularity of $\phi$ then requires that
$\phi(t,0)=\phi(t,L)=0$ as for the commutative case [see
Eq.~(\ref{bstrongc})].
The jump $\Delta\left(\partial_x\phi\right)$ is then proportional to
$\lambda'$, unless $\phi(t,x-y)=-\phi(t,y-x)$ around $y=0$ and
$L$.
This latter condition must strictly hold if one also wants to
take $\lambda'\to\infty$ in the strong coupling limit [otherwise
$\Delta\left(\partial_x\phi\right)$ would diverge].
\par
For finite values of $\lambda'$, we can still use the
potential~(\ref{g}) to estimate the non-commutative corrections by
employing standard perturbation methods.
In particular, we treat the interaction Lagrangian~(\ref{L2}) as a
perturbation with respect to~(\ref{oneplusonelag}) and estimate
the correction to the energy levels of the strongly coupled
($\lambda\to\infty$) unperturbed theory as
\be
\Delta E_n
=\frac{\lambda'}{L\,\ell}\,\int_0^{\ell} \d x\,|\phi_n|^2
\sim
\lambda'\,E_n\,\left(\frac{\ell}{L}\right)^2
\ ,
\label{dE}
\ee
where $\phi_n$ is given in Eq.~(\ref{phi_n}) and the last
approximation holds for $\ell\ll L$.
A similar correction can be computed for the spectrum of
continuous modes outside the plates.
For example, to the left of $x=0$, such modes can be written as
\be
\phi_k(t,x)=\sqrt{\frac{2}{k\,W}}\,e^{-i\,k\,t}\,\sin(k\,x)
\ ,
\ee
where $W$ is a finite interval which must be taken to
infinity at the end of the computation.
The corresponding correction to the energy $E_k=|k|$ is then
\be
\Delta E_k\sim
\lambda'\,E_k\,\frac{\ell^2}{L\,W}
\ ,
\ee
which vanishes in the limit $W\to\infty$ and is thus negligible.
This argument shows that non-commutative boundary corrections
are negligible for continuous spectra, such as is the case of the moving
mirror analysed in Ref.~\cite{mirror}.
\par
>From the above results it follows that the Casimir force between
``fuzzy'' plates is given by
\be
\mathcal{F}_{\rm f.p.}&\!\!\simeq\!\!&
-\frac{\partial}{\partial L}\left[
\sum_n\left(E_n+\Delta E_n\right)
-W\!\!\int \d k\,E_k\right]
\nonumber
\\
&\!\!\sim\!\!&
\mathcal{F}_0\left[
1+3\,\lambda'\,\left(\frac{\ell}{L}\right)^2\right]
\ ,
\label{DF}
\ee
that is
\be
\frac{\mathcal{F}_{\rm f.p.}-\mathcal{F}_0}{\mathcal{F}_{0}}
\sim
\left(\frac{\ell}{L}\right)^2
\ .
\label{DEsE}
\ee
Of course, numerical factors were dropped from the final
expressions since they would just be related to the specific form of the
``fuzzy'' potential~(\ref{g}).
Further, had we considered more spatial dimensions, the difference
would have appeared in an overall (volume) factor which could be
absorbed in the couplings $\lambda$ and $\lambda'$.
\par
A few comments are now in order.
Firstly, the correction in Eq.~(\ref{DF}) explicitly depends on the
coupling constant $\lambda'$ and we have no means of estimating its
value for the (unphysical) case of a scalar field considered here.
One natural assumption would be to set $\lambda'\sim 1$ and,
for the more realistic case of an electromagnetic field, take
$\lambda'\sim \alpha$ (the fine structure constant).
Further, since it is natural to assume that $\ell\ll L$ (otherwise
non-commutativity would have already been discovered in our
macroscopic world), our results will always be given in the perturbative
expansion in powers of $\ell/L$ as we did above.
\subsection{Volume corrections}
\label{NC_v}
Having estimated the effects due to the uncertainty in the
position of the boundaries, we shall now compute volume corrections,
that is the effects due to non-commutativity that remain if we
impose the usual boundary conditions~(\ref{bstrongc}) corresponding
to sharply defined plates.
\par
Using the results reviewed in Section~\ref{NC_st}, we can express
a scalar field in a non-commutative four-dimensional Minkowski
space-time in terms of the field modes as
\be
u_{\rm NC}(t,\vec{x})=
\frac{e^{-\frac{\Theta}{4}\left(\omega^2+p^2\right)}
e^{i\,\vec{p}\cdot\vec{x}-i\,\omega\,t}}{(2\,\pi)^{3/2}\sqrt{2\,\omega}}
\ ,
\ee
where $\vec x=(x,y,z)$, $\omega=-p_0$,
$\vec p=(p_x,p_y,p_z)$ and $p=\sqrt{\vec p\cdot \vec p}$.
The fact that the time is non-commutative is reflected in the equation
above by the presence of the energy $\omega$ in the Gaussian damping
factor, and we recall that the relative sign between terms
proportional to $\omega^2$ and $p^2$ in the exponential is not
affected by the space-time signature.
\par
The scalar field operator in terms of
these field modes takes the form
\begin{widetext}
\be \hat \phi(t,\vec{x})= \int\frac{\d^3
p}{(2\,\pi)^{3/2}\sqrt{2\,\omega}} \left[\hat b_{\vec
p}\,e^{-\frac{\Theta}{4}\left(\omega^2+p^2\right)}
e^{i\,\vec{p}\cdot\vec{x}-i\,\omega\,t} +\hat b_{\vec
p}^{\dag}\,e^{-\frac{\Theta}{4}\left(\omega^2+p^2\right)}
e^{-i\,\vec{p}\cdot\vec{x}+i\,\omega\,t}\right], \ee and the
momentum density becomes \be \hat \pi(t,\vec{x})= \int\frac{\d^3 p
\left(-i\omega\right)}{(2\,\pi)^{3/2}\sqrt{2\,\omega}}\,
\left[\hat b_{\vec p}\,e^{-\frac{\Theta}{4}
\left(\omega^2+p^2\right)} e^{i\,\vec{p}\cdot\vec{x}-i\,\omega\,t}
-\hat b_{\vec p}^{\dag}\,
e^{-\frac{\Theta}{4}\left(\omega^2+p^2\right)}
e^{-i\,\vec{p}\cdot\vec{x}+i\,\omega\,t} \right] \ , \ee
\end{widetext}
with the ladder operators satisfying
\be
\left[\hat b_{\vec p},\hat b_{{\vec p}'}^{\dag}\right]
=\delta^{(3)}(\vec p-\vec p')
\ .
\ee
\par
The massless free-field Hamiltonian
\be
H=\frac{1}{2} \int \d^3 x
\left[\pi^2+\left(\vec\nabla\phi\right)^2\right]
\ ,
\ee
then becomes
\be
\hat H= \int \frac{\d^3 p}{(2\,\pi)^3}\,\omega(p)\,
e^{-{\Theta}\,\omega^2}
\left(\hat b_{{\vec p}}^{\dag}\,\hat b_{\vec p}+\frac{1}{2}\right)
\ ,
\ee
and the zero point energy
\be
E_{\rm ZP}&\!\equiv\!& \langle 0|H|0\rangle
\nonumber
\\
&\!=\!&\int_0^\infty
\frac{p^2\,\d p}{(2\,\pi)^2}\,
e^{-\Theta\,p^2}
\omega(p)
\ ,
\label{ZPE}
\ee
which is finite since the Gaussian damping factor dominates
at large $p$ and makes the above integral converge.
\par
Let us assume that the two perfectly conducting plates are located
at $x=0$ and $L$ and extend infinitely in the $(y,z)$-plane.
The boundary conditions~(\ref{bstrongc}) then imply that, between the
plates, the momentum along $x$ is quantized,
\be
p_x = \frac{\pi\,n}{L},
\quad
n\in Z
\ ,
\ee
whereas $\vec{k}=(p_y ,p_z)$ takes continuous values.
The frequency of a massless scalar field
mode can therefore be written as
\be
\omega_{n}(k)
=\sqrt{\frac{\pi^2\,n^2}{L^2} +k^2}
\ ,
\label{omega}
\ee
with
$k=\sqrt{\vec k\cdot\vec k}$.
\par
We can rewrite the energy density from Eq.~(\ref{ZPE}) in the
volume contained between the plates using Eq.~(\ref{omega})
as
\be
E_{\rm in}&=&\frac{\mathcal{A}}{4\,\pi}\,
{\sum_{n=-\infty}^{\infty}} \int_{0}^{\infty} k\,\d k\,
e^{-\Theta\,\omega^2_{n}(k)} \omega_{n}(k)
\nonumber
\\
&=&\frac{\mathcal{A}}{2\,\pi}\, {\sum_{n=0}^{\infty}}'
\int_{0}^{\infty} k\,\d k\, e^{-\Theta\,\omega^2_{n}(k)}
\omega_{n}(k)
\ ,
\label{energy0}
\ee
where $\mathcal{A}$ is the area of the plates and the prime on the
summation in Eq.~(\ref{energy0}) means that the term with $n=0$ is
weighted by a factor of 1/2.
Since the plates extend along $x$ and $y$ infinitely, the sums over
the parallel wave vectors are replaced by integrals.
We then change the integration variable from $k$ to $\omega$ and
use~(\ref{omega}) to write the energy density as
\be
E_{\rm in}
=\frac{\mathcal{A}}{2\,\pi}{\sum_{n=0}^{\infty}}'
\int_{\frac{\pi\,n}{L}}^{\infty} \d\omega\, e^{-\Theta\,\omega^2}
\omega^2
\ .
\label{energy_in}
\ee
\par
The energy density outside the plates can also be obtained by
noting that the momentum along the $x$ direction takes continuous
values there and in this region the summation over $n$ from the
previous equation becomes an integral,
\be
E_{\rm out}
=\frac{\mathcal{A}}{2\,\pi}{\int_{0}^{\infty}} \d n\,
\int_{\frac{\pi\,n}{L}}^{\infty} \d\omega\, e^{-\Theta\,\omega^2}
\omega^2
\ .
\label{energy_out}
\ee
We again remark that the
expressions for the energy density in the previous equations are
finite and do not need to be regularized.
\begin{figure*}[t!]
\centering{
\raisebox{3cm}{$\mathcal{F}$}
\includegraphics[width=0.45\textwidth]{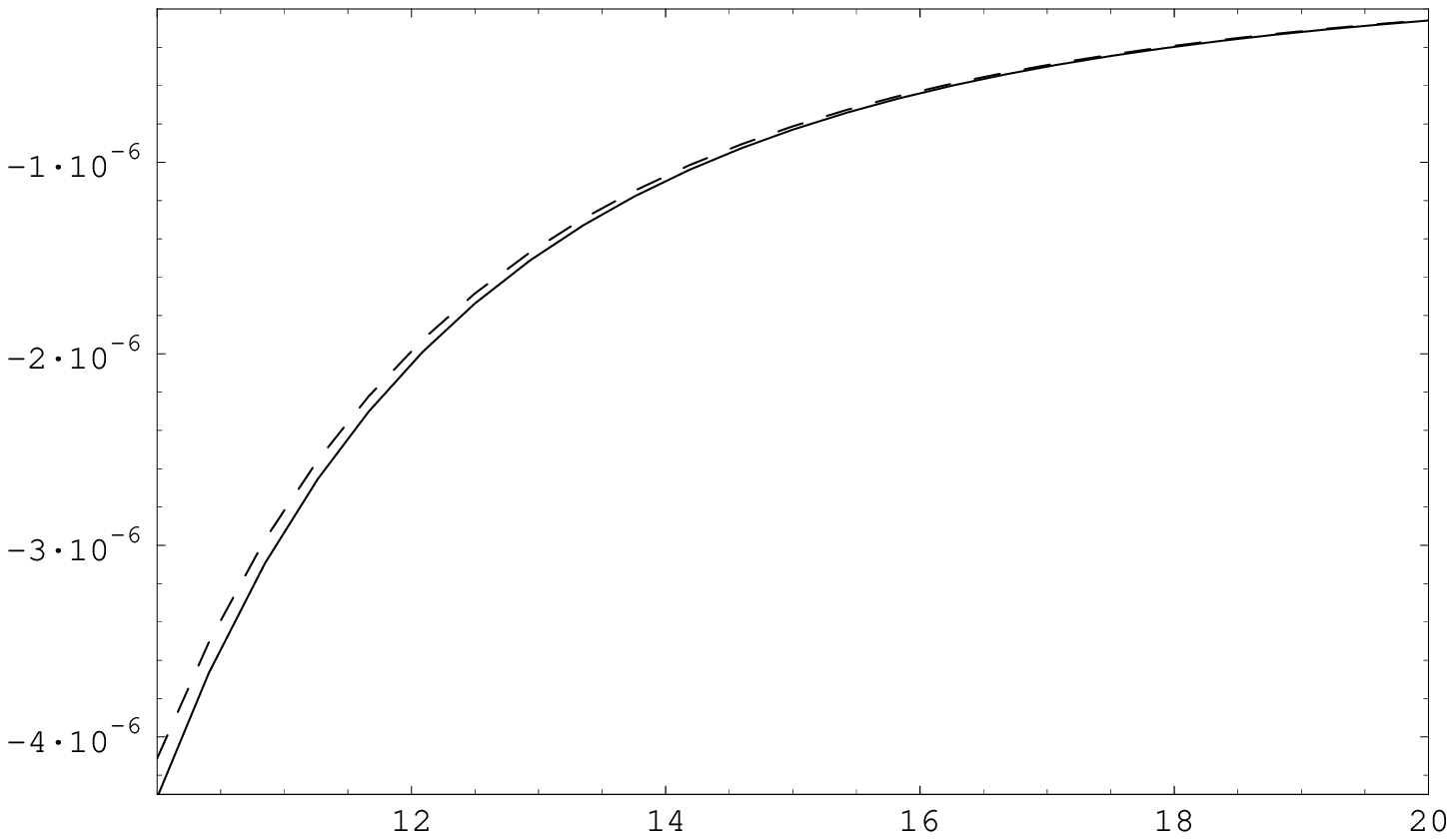}
$\ $
\raisebox{3cm}{$\frac{\Delta \mathcal{F}}{\mathcal{F}}\%$}
\includegraphics[width=0.45\textwidth]{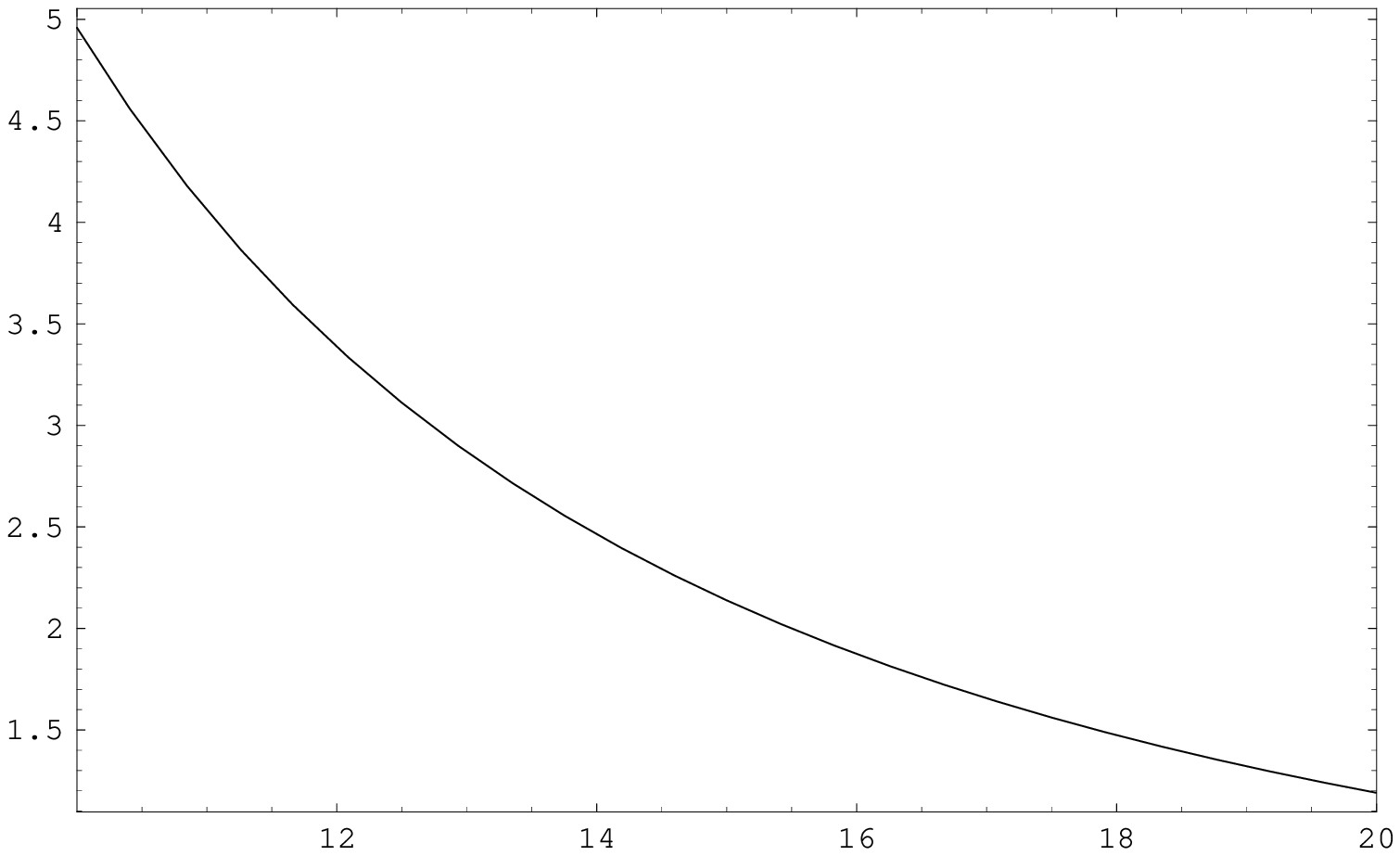}
\\
\hspace{1cm}$L/\ell$
\hspace{9cm}$L/\ell$
}
\caption{Left plot: Casimir pressure in the commutative (dashed line) and
non-commutative case (solid line) in units of $\Theta^{-4}$
as a function of the separation $L$
between the plates (in units of $\ell=\sqrt{\Theta}$).
Right plot: Percentage difference between the two Casimir pressures
as a function of $L/\ell$.
Only volume effects are included.
\label{F4} }
\end{figure*}
\par
After performing the integration over $\omega$ in Eq.~(\ref{energy_in}),
we can calculate the force acting on the plates from inside as
(recall that $\Theta=\ell^2$)
\be
F_{\rm in}=
-\frac{\partial E_{\rm in}}{\partial L}
=-\frac{\mathcal{A}\,\pi^2}{2\,L^4}
{\sum_{n=0}^\infty}'\,
n^3\,e^{-\frac{\ell^2}{L^2}\,\pi^2\,n^2}
\ ,
\label{force_in}
\ee
and a similar expression is obtained for the force $F_{\rm out}$ acting
on the plates in the outer region with the summation over
$n$ again replaced by an integration.
The net Casimir force is then found by subtracting the force acting
on one side of the plate from the force acting on the other side,
\begin{widetext}
\be
F_{\rm net}=F_{\rm in}-F_{\rm out}
=
-\frac{\mathcal{A}\,\pi^2}{2L^4}
\left[
{\sum_{n=0}^{\infty}}'\,n^3\,e^{-\frac{\ell^2}{L^2}\,\pi^2\,n^2}
-\int_{0}^{\infty}
\d n\,n^3\,e^{-\frac{\ell^2}{L^2}\,\pi^2\,n^2}
\right]
\ .
\ee
We use the Euler-Maclaurin formula,
\be
\sum_{n=1}^N\,f(n)-\int_0^N\d n\,f(n)=
\frac{f(0)-f(N)}{2}
+\sum_{k=1}^\infty
\frac{B_{2\,k}}{\left(2\,k\right)!}\,
\left[f^{(2\,k-1)}(N)-f^{(2\,k-1)}(0)\right]
\ ,
\ee
\end{widetext}
where $B_{2\,k}$ are Bernoulli's numbers,
and the fact that in our case
\be
f(0)=\lim_{N\to \infty}f(N)=
\lim_{N\to \infty} f^{(2\,k-1)}(N)=0
\ ,
\ee
to finally obtain the pressure as a power series in
$\ell/L$,
\be
\!\!
\mathcal{F}
&\!\!=\!\!&
\frac{F_{\rm net}}{\mathcal{A}}
\label{dF_v}
\\
&\!\!=\!\!& \mathcal{F}_0 \left[
1
+\frac{17\,\pi^2}{21}\left(\frac{\ell}{L}\right)^2
+\frac{\pi^4}{4}\left(\frac{\ell}{L}\right)^4 +\dots \right]
\ .
\nonumber
\ee
The factor $\mathcal{F}_0$ is Casimir's original
result~(\ref{Ccom}), and for a comparison with the
complete result including 100 orders in $\ell/L$ see
Fig.~\ref{F4}.
For a plate separation $L\simeq 10\,\ell$ the volume
correction is of the order of $5\%$ and decreases rapidly to zero
for larger $L$.
\par
We conclude this Section by noting that the leading order of the
volume correction is the same $(\ell/L)^2$
as that of the boundary effects estimated in Section~\ref{NC_b}
and, if one regards $\ell$ as the ``fundamental'' plate thickness,
both the boundary correction~(\ref{DEsE}) and the leading order
volume correction~(\ref{dF_v}) have the same functional dependence
on $\ell/L$ as finite surface effects.
\section{Comparison with surface effects}
\label{surf}
When evaluating the Casimir force for real media, one should
consider temperature corrections, corrections which appear because
of the finite conductivity of the plates and because of the roughness
of the plates.
As we mentioned at the end of the previous Section,
the latter compare very well to the type of modifications which are
induced by space-time non-commutativity and will now be briefly
reviewed (see, {\em e.g.}, Refs.~\cite{bordag,milton}).
\par
The boundaries for real plates, no matter which geometry is under
discussion, show deviations from perfect figures.
For example, the plates can be slightly off from being parallel to each
other, or the surfaces of the plates have some imperfections.
The surface roughness of any real material contributes to the magnitude
of the Casimir force.
In any of these cases, the amplitude of these
distortions $A$ is much smaller than the separation of the two
plates $L$.
Even so, for plate separations of the order of $L\sim 1\,\mu$m,
these contributions are rather large and need to be
taken into consideration when comparing theoretical models
with experimental data.
\par
Let us first consider the Casimir force between two square parallel
plates with sides of finite length $d$.
For the derivation to be valid (we considered the
case of two infinite plates) we need to have $d\gg L$.
We further assume that instead of the plates being exactly parallel,
they are placed with an angle $\alpha$ between them.
We want $|\alpha|\,d\ll L$ so this angle needs to be very small
(this is one of the examples of deviations from the parallel plane
geometry).
The corrections up to the fourth order in $|\alpha|\,d\ll L$ to the
Casimir force in this case are given by~\cite{bordag}
\be
\mathcal{F}^{\alpha}=\mathcal{F}_0
\left[1+\frac{10}{3}\left(\frac{\alpha\,d}{L}\right)^2
+7\left(\frac{\alpha\,d}{L}\right)^4\right]
\ ,
\label{dF_d}
\ee
where
$\mathcal{F}_0$ is again the expression in Eq.~(\ref{Ccom}).
\par
For plates covered by short scale roughness, let us describe the
location of the surfaces of the two plates by
\be
\begin{array}{l}
x_1=A_{1}\,f_{1}(y_{1},z_{1})
\\
\\
x_2=L+A_{2}\,f_{2}(y_{2},z_{2})
\ .
\end{array}
\ee
where $L$ is now the mean distance between the two plates.
The values of the roughness amplitudes are chosen such that
max$|f_{i}(y_{i},z_{i})|=1$.
The mean values of the location of the plates on the $x$-axis are
given by
\be
\expec{x_1}\equiv
A_{1}\,\expec{ f_{1}(y_{1},z_{1})}=0
\ee
and
\be
\expec{x_2}\equiv
L+A_{2}\,\expec{f_{2}(y_{2},z_{2})}=L
\ .
\ee
We assumed here that
the amplitudes $A_1$ and $A_2$ are much smaller than the mean
distance between the plates $L$, and $L$ is much smaller than both
the size of the plates $d$ and their thickness $t$.
At the same time in all real situations we have $L/t$, $L/d\ll |A_i|/L$
so that we are looking for a perturbation expansion in powers of $|A_i|/L$
to zero order in $L/t$ and $L/d$. For this case the Casimir force
takes the following form \cite{bordag}
\begin{widetext}
\be
\mathcal{F}^{\rm R}&\!=\!&
\mathcal{F}_0
\left\{1+10
\left[\expec{f_1^2}\left(\frac{A_1}{L}\right)^2
+\expec{f_2^2}\left(\frac{A_2}{L}\right)^2\right]
\right.
\nonumber
\\
&&
\phantom{\mathcal{F}_0}\
\left.
+35\left[
\expec{f_1^4}
\left(\frac{A_1}{L}\right)^4
+6\,\expec{f_1^2\,f_2^2}
\left(\frac{A_1}{L}\right)^2\left(\frac{A_2}{L}\right)^2
+\expec{f_2^4}\left(\frac{A_2}{L}\right)^4\right]
\right\}
\ .
\label{dF_r}
\ee
\end{widetext}
\par
In the previous sections we calculated the leading order (in $\ell/L$)
corrections to the Casimir force due to the presence of
boundaries and the volume corrections up to $(\ell/L)^{100}$ in a
non-commutative scenario.
We now see that, at least at leading order, the corrections due
to space-time non-commutativity have the same sign and
a similar dependence on $1/L^2$ as those due to the imperfections
of the plates (orientation and roughness).
To be more accurate, the corrections~(\ref{dF_d})~and~(\ref{dF_r})
due to physical imperfections of the plates depend on the square
of the ratio between the amplitude of the functions which describe
the deviations from perfectly flat plates and the plate separation.
In the non-commutative case there are two types of corrections,
one due to the ``fuzziness'' of the plates [estimated in Eq.~(\ref{DF})]
and the other due to non-commutativity of space-time
coordinates in the volume between the plates [see Eq.~(\ref{dF_v})].
We have shown that both of them are proportional to $\Theta/L^2$,
where $\Theta=\ell^2$ is the parameter which describes
non-commutativity, which can therefore be also viewed as describing
an intrinsic (minimum) plate thickness.
\section{Conclusions}
\label{conc}
In the present work we calculated non-commutative corrections to
the scalar Casimir force between two parallel plates. We used a
coherent states approach to non-commutativity first introduced by
Smailagic and Spallucci in Refs.~\cite{smailagic}, which produces
results that do not depend on the self-interaction of the scalar
field.
\par
There are two types of corrections to the attractive force between
the plates.
The first type of correction is what we called a ``boundary'' correction.
Defining a boundary in the classical sense in a non-commutative
scenario is very problematic.
We therefore took a heuristic approach to the problem and defined a
``fuzzy'' boundary which we modelled using a potential different
from zero in an interval of the order of the fundamental length
introduced by non-commutativity.
Treating this term as a perturbation to the part of the Lagrangian
which describes the interaction between the scalar field and the plates,
we were able to estimate corrections which appear because of the
non-commutativity of the boundaries.
As a result, the scalar Casimir force is increased.
We have also shown that in the limit in which the boundaries
become sharp, there will still be a second contribution from
non-commutativity in the volume between the plates.
The non-commutativity of space-time in fact suppresses the
plane waves by a Gaussian factor and thus affects the vacuum
energy density.
This latter contribution also has the same sign as
the commutative part of the Casimir force, and we can therefore
conclude that non-commutativity is in general expected to increase
the (scalar and attractive) force between two parallel plates.
\par
The treatment of non-commutative geometries with ``boundaries'' is a
long-standing problem.
The phenomenological approach we have
developed here may be of use for other physical theories which
involve non-commutative spaces with boundaries since it allows
explicit calculations of contributions due to non-commuting
aspects of space-time.
\end{document}